# On the anisotropy of stress-distribution induced in glasses and crystals by non-ablative femtosecond laser exposure

Ben McMillen,[*] Yves Bellouard

*Department of Mechanical Engineering,*
*Eindhoven University of Technology, Den Dolech 2, 5612AZ Eindhoven*
*The Netherlands*
[*]*ben@mcmillen.eu*

**Abstract:** Femtosecond laser exposure in the non-ablative regime induces a variety of bulk structural modifications, in which anisotropy may depend on the polarization of the writing beam. In this work, we investigate the correlation between polarization state and stress anisotropy. In particular, we introduce a methodology that allows for rapid analysis and visualization of laser-induced stress anisotropy in glasses and crystals. Using radial and azimuthal polarization, we also demonstrate stress states that are nearly isotropic.

## 1. Introduction

The use of ultrafast lasers presents a new paradigm of optical fabrication, owing to the precise control of highly localized material modifications through nonlinear absorption processes. This fabrication technique has enabled the realization of 2- and 3D structures in bulk dielectrics [1-12]. Classifications of laser-induced structural changes in dielectrics generally fall into one of three categories: continuous modifications [3,6,8,9,13], self-organized nanostructure formation [12,14-19], and nano-void formation [20-22]. These modifications may induce various types of anisotropy in the material that can have strong effects on the chemical selectivity [23], optical [24], thermal [25], and mechanical properties [26]. For instance, in silica, it was reported that structural transformations are accompanied by a volume change [21,23,26,27] which induces stress in the surrounding substrate [28,29]. This stress can be either tensile (first regime, continuous modifications [26]) or compressive (second regime, nanostructure formation [29]), where stress depends on the orientation of nanograting formation [29].

Laser-induced anisotropy plays a key role in the stress state of laser-machined devices, and can have drastic effects on both the quality and speed of manufacture. In some cases, stress may be desirable, as it locally influences etching rate [30,31], potentially leading to faster processing times. However, for optical applications such as laser-written waveguides, localized stress increases birefringence [17,26,32-36], and can also lead to cracking in guiding structures [37,38]. These undesirable effects contribute to increased losses, reducing the overall quality and performance in demanding applications. Given these examples, it is clear that more information about the stress state of laser-machined structures and the associated anisotropy would greatly benefit the process of laser-fabrication.

Anisotropy in the laser-writing process may find its origin from various sources, including inhomogeneity of the laser beam [39], pulse front tilt [40], material anisotropy [41], or the formation self-organized nanostructures with preferential orientation [14,16]. In order to identify the nature and origin of this anisotropy, it is necessary to decouple each of the possible contributions.

In this work, we investigate the use of highly symmetric writing patterns as an analysis tool for identifying various stress-related contributions to anisotropy in the writing process. Specifically, these patterns induce a measureable stress-induced birefringence, providing an indication of the source of the anisotropy, as well as its dependence on laser exposure parameters. This birefringence is then correlated with the geometry and orientation of the associated nanostructure as revealed through scanning electron microscopy. We further show that this methodology can be applied to a variety of transparent materials.

## 2. Methodology

Decoupling contributions to anisotropy in the laser-writing process presents some interesting challenges, and imposes limits on the choice of writing geometry. Ideally, we require a pattern that is symmetric and does not introduce any directionally dependent effects, which may further influence the measurements. Due to these limitations, and for simplicity, we have chosen cylindrical structures, whose shape is inherently symmetric in the plane of writing.

In practice, these structures are composed of individual rings, stacked with a fixed spacing to form a tube. Each ring is formed by focusing light from an ultrafast laser system (Amplitude Systèmes, 1030 nm, 275 fs, 400 kHz) into the bulk of a silica substrate with a 20x (0.4 NA) objective. The substrate, which is mounted to a set of high-bandwidth (~4 kHz) piezoelectric flexure stages, is then translated in a circular fashion (see Fig. 1a), forming a ring. Rings are then stacked by translating the focus a fixed distance along the $z$-direction (~5 µm) to form a tube, and the process is repeated (see Fig. 1b). The dimensions of an individual ring are defined by the size of the laser-modified zone (~2 µm x 8 µm) so that some overlap between adjacent rings is present. We call this structure a 'stressor'.

After fabrication, false color intensity retardance maps of each tube structure are generated with a standard microscope (Olympus BX51) fitted with a commercial system to measure optical retardance (CRi-Instruments). Stress measurement is facilitated through photoelasticity, which gives a direct measurement of the induced stress in the substrate [42].

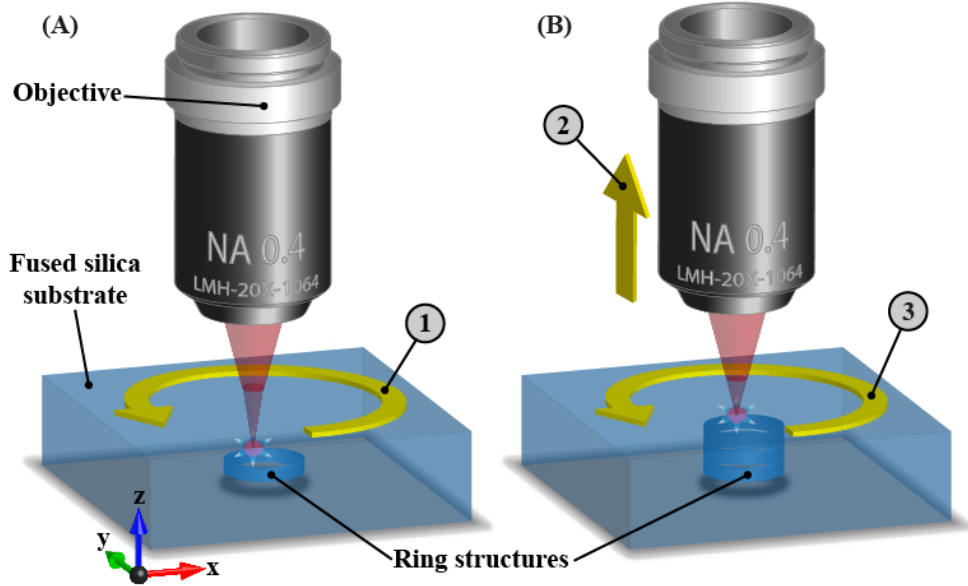

Fig. 1. Schematic illustration of the process used for fabricating tube-based stressors. A tube is composed of individual rings (A), fabricated by translating the substrate in a circular motion (1) at a fixed velocity. Rings are then stacked (B) by translating the focus along the z-direction (2) and the process repeated (3), forming a tube. Focal-plane separation (z-distance between rings) was fixed at 5 um.

The measured retardance of these patterns is directly related to the induced stress through a local change in the difference of refractive index. The refractive index in a dielectric medium can be represented by an ellipsoid known as the indicatrix [42], which is given by:

$$B^o \left( x_1^2 + x_2^2 + x_3^2 \right) = 1 \qquad (1)$$

where the (un-perturbed) refractive index is found through the relation $B^o = 1/(n_o)^2$. In fused silica, which is an isotropic material, this surface is a perfect sphere, and the refractive index is the same regardless of direction. The quantity $B^o$ may be related to distortions of the ellipsoid (and the corresponding changes in refractive index) due to the application of stress or an electric field [42], as follows:

$$\Delta B_{ij} = z_{ijk} E_k + \pi_{ijkl} \sigma_j \qquad (2)$$

$E_k$ is the electric field, $\sigma_j$ the applied stress, and the coefficients $z_{ijk}$ and $\pi_{ijkl}$ represent third and fourth rank tensors for the electro-optical and piezo-optical coefficients respectively. For our purposes, we only require the portion involving stress. The 4$^{th}$ rank tensor $\pi_{ijkl}$, which specifies the piezo-optical coefficients, gives information about the magnitude and orientation of the stress field. Polarization microscopy measures these two quantities in the form of optical retardance. The difference of principal stresses may be calculated from the measured retardance using the following relation [42]:

$$\sigma_1 - \sigma_2 = \frac{R}{T(C_1 - C_2)} \tag{3}$$

where the constant $C_1$-$C_2$=3.55x10$^{-12}$ Pa and is related to the piezo-optical coefficients for fused silica [43] by:

$$C = \left(\frac{n^3}{2}\right)(\pi_{11} - \pi_{12}) \tag{4}$$

For the purposes of this study, we are focused mainly on the *intensity* portion of the retardance.

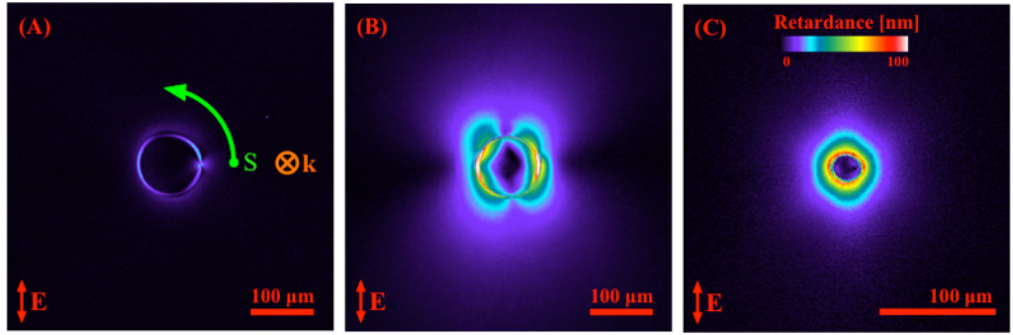

Fig. 2. Illustrative examples of laser-induced stress anisotropy in (a) borosilicate glass, (b) fused silica, and (c) crystalline quartz. Although the three patterns have similar geometry, written under the same polarization and focusing conditions (but with different laser fluence adapted for each material), the stress-induced retardance profiles are noticeably different. These simple examples demonstrate how stress induced birefringence resulting from symmetric laser-written patterns may be used to reveal anisotropy in the laser-writing process. This anisotropy may be due to the presence of self-organized nanostructure, anisotropy present the writing beam, or the structure of the material itself. In (a), the green arrow indicates the starting position and direction of writing for each successive tube layer (for all materials). The polarization of the writing beam was linear, and fixed parallel to the vertical axis of the figure.

Fig. 2 shows example retardance maps of laser-written tubular structures in three different materials (borosilicate glass, fused silica, and crystalline quartz), illustrating how these simple structures can be applied to reveal anisotropy in the laser-writing process. Striking differences are immediately apparent, despite identical geometry, writing polarization, and focusing conditions (but with laser fluence adapted for each material). Fig. 2a (borosilicate) represents a case where the laser-writing process introduces almost no anisotropy, producing a pattern that is symmetric and homogeneous. This is in contrast to Fig. 2b (fused silica), which suggests anisotropy in the laser pattern itself, resulting in a highly anisotropic retardance map. Finally, the retardance image displayed in Fig. 2c (crystalline $SiO_2$) hints at anisotropy due to the structure of the material.

In the next section, we discuss how these patterns can be analyzed to discern various sources of anisotropy in amorphous and crystalline $SiO_2$.

## 3. Experimental results

### 3.1 Fused silica

We focus our attention on fused silica due to its practical importance, as well as its relevance to the fabrication of devices for various applications [2,4,6,30,44,45]. In the manufacture of such devices, anisotropy induced in the substrate during laser processing can lead to adverse effects, which ultimately hinder device performance and increase manufacturing times (as in the case of wet-etch processes). Often, these effects are due to the presence of self-organized nanostructures (i.e. 'nanogratings') within the laser-modified zone, which are known to form perpendicular to the writing polarization [14].

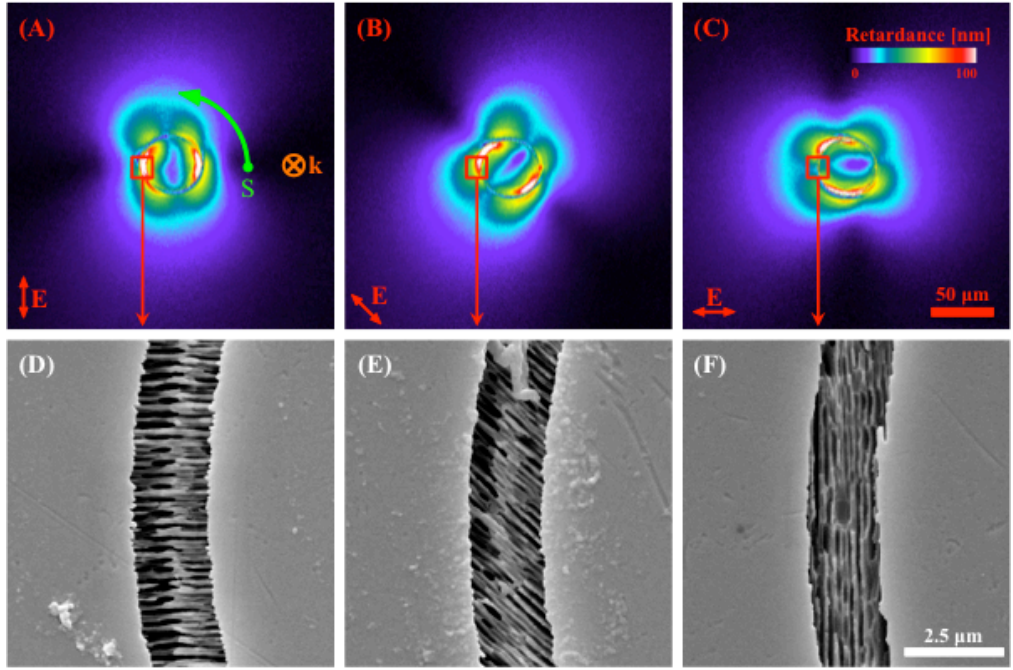

Fig. 3. – Demonstration of the dependence of anisotropic stress field orientation on writing polarization. Retardance maps (a-c) and corresponding SEM images (d-f) of nanograting structure are shown for 50-μm-diameter tubes fabricated with linear polarization orientations of 90°, 45°, and 0° respectively. The green arrows indicate the starting position and writing direction for successive layers in each tube. The pulse energy was with a writing speed of 1 mm/s and a fixed $z$-spacing of 5 μm.

In the case of tube structures, if the induced stress-anisotropy is due to the formation of nanogratings, then rotation of the (linear) writing polarization with respect to a fixed sample orientation should yield a change in the orientation of the anisotropic retardance field. To verify this hypothesis, a series of tubes with a diameter of 50 μm and $z$-spacing of 5 μm, was written through the entire thickness of a 25 x 25 x 1 mm silica substrate using linear polarization. Between successive tubes, the polarization was rotated between 90°, 45°, and 0° degrees. The as-fabricated structures were then imaged to obtain retardance maps, followed by a 5 minute etch in 2.5% hydrofluoric acid to reveal the nanograting structure in preparation for imaging with a scanning electron microscope (SEM). The results of this experiment are shown in Fig. 3.

Anisotropy can also arise from the choice of writing direction [40], and can be revealed simply by writing in the opposite direction. In the case of tube structures however, this is not entirely effective. Tube patterns are already axially symmetric, meaning that directionality in writing will have little (if any) effect. This is further compounded by the fact that nanograting induced retardance will be of much greater magnitude than any possible contributions from directional effects alone, as stresses generated between grating lamella can approach several GPa [27]. Therefore, any subtle changes arising from writing directionality will be difficult to detect.

Instead, we propose the use of symmetric polarization states that are likely to produce directionally *independent* nanostructures. Similar to the linear polarization case (see Fig. 3), a series of tubes was written, first using circular polarization as a reference, followed by radial and azimuthal states, which were generated using a passive polarization converter [44].

Writing conditions were identical to the linear polarization case, with the exception of radial and azimuthal, which required a pulse energy of 1 µJ, as these states produce a larger spot size at the objective focus. The retardance maps and corresponding nanograting structure for these states are shown in Fig. 4.

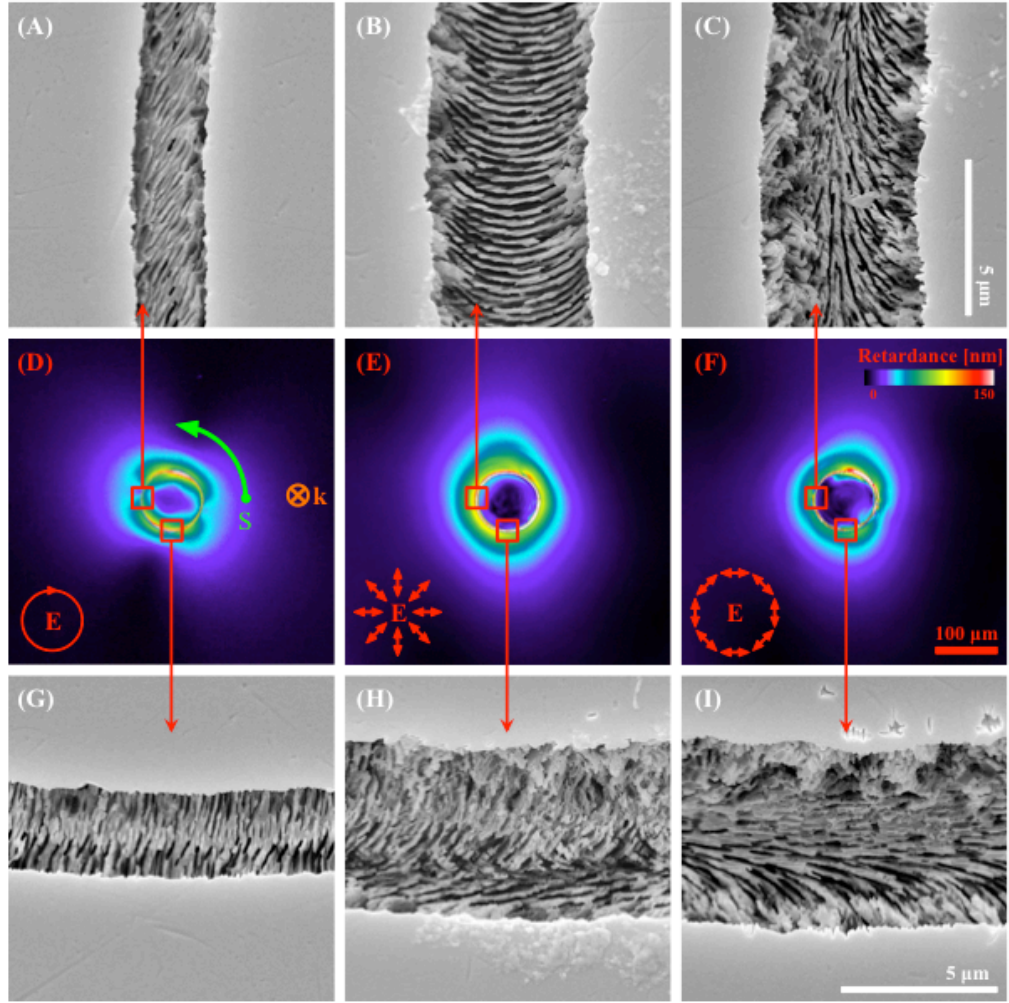

Fig. 4. Examples of isotropic retardance plots for (d) circular, (e) radial, and (f) azimuthal polarization. Circular polarization induces an anisotropic stress, while radial and azimuthal are nearly isotropic. Irregularities are found in the nanograting formation as shown in the corresponding SEM images above and below each field map (a-c, g-i). Writing speed was fixed at 1 mm/s with a pulse energy of 200 nJ for circular polarization and 1 uJ for radial and azimuthal polarization.

We begin by comparing the retardance map for circular polarization (d) with those of radial (e) and azimuthal (f). Although counterintuitive at first, we find that circular polarization induces a retardance field that is highly anisotropic, and closely resembles the retardance maps obtained for linear polarization. This is reflected in the structure of the nanogratings (a, g), which are nearly linear, and appear to be a mixing between two orthogonal (linear) states. These 'hybrid' lamellae also have a preferential orientation, which closely mimics the orientation of the fast axis of the wave plate (45°). We attribute this observation to several factors, including the difficulty in obtaining a 'pure' circularly polarized beam, and on other factors such as Fresnel reflections at the air-glass interface.

We now focus our attention on the radial and azimuthal polarization case (e, f). Here we find retardance maps that appear nearly isotropic, as predicted. Small distortions are found at the 'poles' of the retardance field, suggesting irregularities in the underlying structure. At first, the grating structures appear regular and well formed (b, c), closely following the orientation of the electric field vector for each polarization case. However, closer examination of the 'polar' regions near the bottom of each ring reveals inconsistencies in the grating lamellae (h, i). Similar to the case for circular polarization, formations appear as a mixing of states. It is likely that the loading in this region is uneven due to the varied grating structure, leading to the distortions found in the retardance maps. This might suggest irregularities in the beam itself, however we have yet to formulate an explanation for this phenomenon.

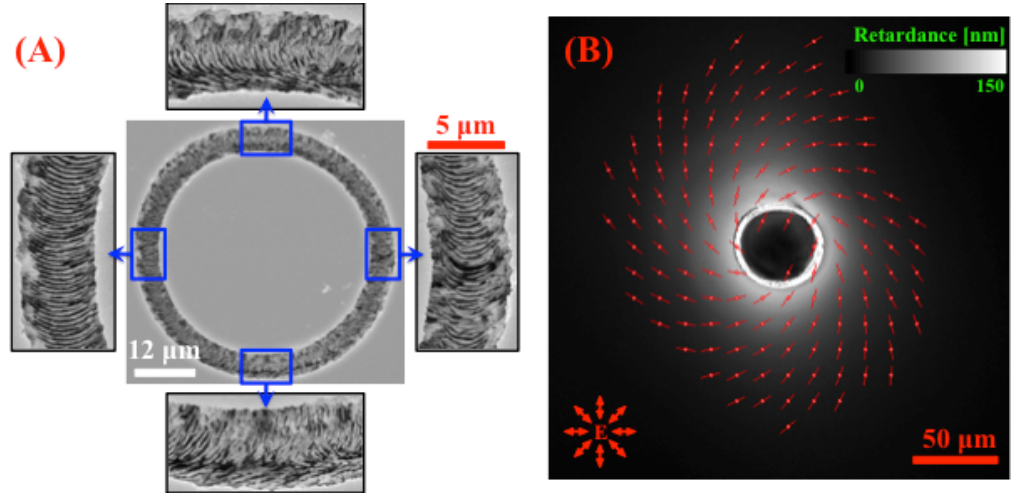

Fig. 5. Nanograting formation in the case of radial polarization manifests asymmetric (anisotropic) geometries that are not entirely apparent from the intensity of the retardance alone, which appears to be nearly isotropic. The grating structure itself, though well formed, has a preferential orientation (concave-up) around the radius of each tube, and is mirrored left-to-right (a). Additional irregularities are also present at the 'poles' of the feature. The stress produced by this configuration is oriented tangentially to the tube structure, generating a retardance pattern whose slow-axis orientation is also anisotropic (b).

It should be noted that, despite using so-called symmetric polarization states, some forms of anisotropy are still present. As shown in Fig. 5, this is especially evident in the structure and organization of the nanogratings for radial and azimuthal polarization. Intuitively, we might expect that the structure shown in Fig. 4b is found with a preferential orientation around the entirety of the ring, and follows the direction of travel of the writing beam (counter-clockwise) during fabrication. As shown in Fig. 5a, this is not the case. The orientation (convex down) is the *same* on both sides.

Anisotropy is also present in the retardance maps. For the purposes of measuring the orientation and distribution of the retardance field, we have used *only* the intensity of the retardance, while leaving out information about the orientation of the slow axis. Fig. 5b shows this information overlaid onto the intensity retardance map of the feature shown in (a). While it is evident that the slow-axis orientation not symmetric, further information about the nature of the stress can be gained.

This orientation anomaly may be explained if we consider that deformations generated by the nanograting lamella produce a maximum stress component oriented perpendicular to the each lamella. The lamellae themselves have a peculiar shape, resulting from the use of radial polarization during writing, and are preferentially oriented in a semi-continuous fashion around the tube. It follows then that, due to the grating shape and orientation, the component of maximum stress will be oriented tangentially to the outer diameter of the structure. This is reflected in the orientation of the slow axis, which corresponds to the directional component with the largest refractive index, as shown in Fig. 5b.

So far, we have only considered the case for isotropic materials; however this method can be applied to anisotropic (crystalline) materials as well, revealing not only anisotropy due to laser writing, but also anisotropy of the material itself.

*3.2 Crystalline Quartz (c-SiO$_2$)*

Tubes were written through both surfaces of a 125 μm thick, z-cut crystalline quartz substrate using similar conditions as previous examples in amorphous silica (400 kHz, 1 mm/s, 0.4NA, 5 μm *z*-spacing) using both rotated linear (90°, 45°, 0°, 200 nJ) and radial (1 μJ) polarization. The x-axis (crystallographic axis) of the substrate was oriented parallel to the *y*-axis of the motion stage (*y*-axis of the figure). To aid visualization of anisotropy in the retardance patterns, the tube diameter was reduced to 25 μm, as shown in Fig. 6.

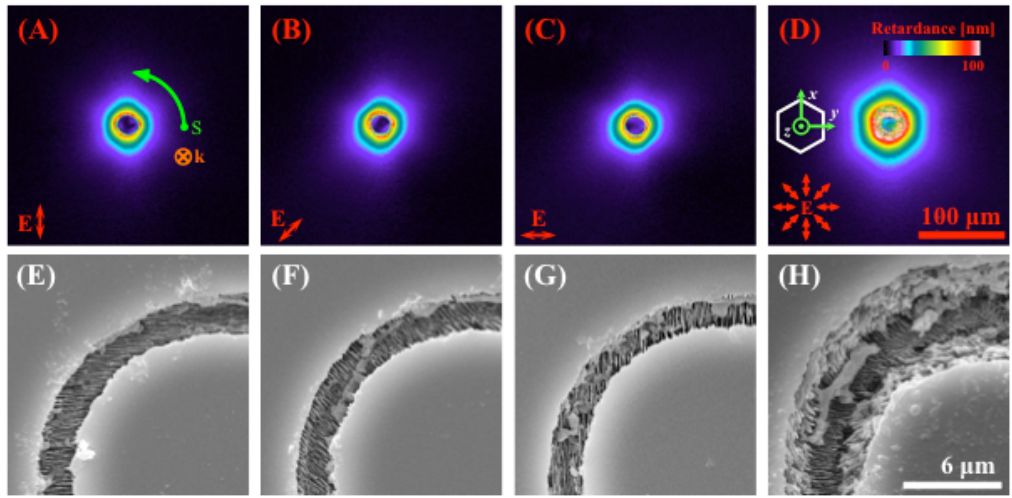

Fig. 6. – Retardance images for 25-μm diameter tubes written in crystalline quartz with linear writing polarizations of 90, 45, and 0 degrees respectively (a-c). In (d), a tube written with radial polarization is given as a comparative example of a stress state that is free of anisotropy induced by directionally dependent nanostructure formation. For the linearly polarized features (a-c), note the visible deformation of the stress field, which is most visible in (b), confirming the formation of laser-induced nanostructure formation as shown in the SEM images (e-h). The triad in (d) indicates the orientation of the crystallographic axes of the substrate.

The retardance patterns displayed in Fig. 6a-d reveal a stress-field that closely mimics the structure of the crystal lattice. This is most evident for the case of radial polarization (d), which generates a retardance map that is nearly free of distortion. The retardance maps for linear polarization appear similar (a-c), with the exception of small distortions, corresponding to the polarization orientation. Similar to silica, nanogratings are visible in the etched patterns (e-h), and are consistently oriented perpendicular to the polarization of the writing beam. For radial polarization, the trend is also similar, with the exception of a change in the periodicity of the nanostructure.

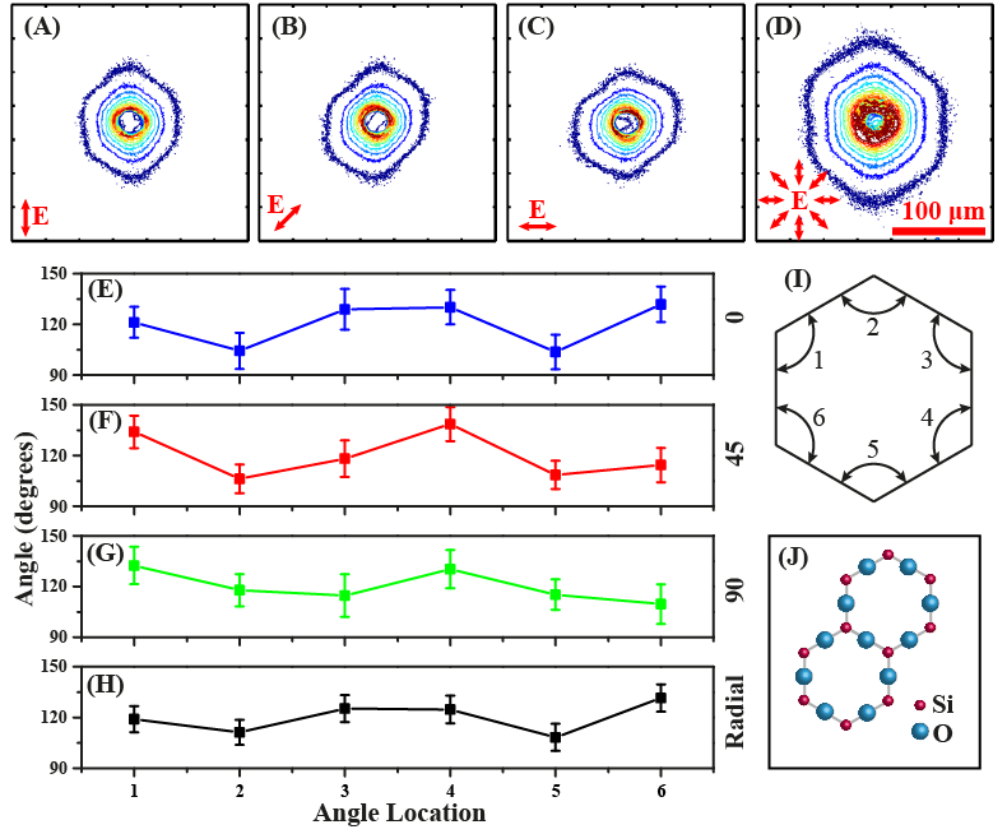

Fig. 7. Quantitative analysis of the lattice distortion in β-quartz due to anisotropic stress. In (a-d), the intensity retardance maps displayed in Fig. 6 have been processed to produce a set of contours at fixed values of retardance. The largest, lowest-value contour (shown in dark-blue) was extracted for analysis. The plots shown in (e-h) represent the angles between a linear fit of each of the facet point groups shown in (a-d) for each respective writing polarization. The reference location of each angle in relation to the hexagonal contour is given in (i). For reference, the structure of β-quartz is also displayed in (j). This method of analysis clearly reveals the presence of anisotropic stress due to the formation of nanogratings within the laser-modified zone. As with the previous analysis in fused silica, radial polarization is used as a reference case for isotropic stress. This is reflected in the plot shown in (h), where the angles between each face are approximately equal. At locations 2 and 5, the angle is slightly smaller. This may be attributed to two factors: the slightly anisotropic distribution of stress at the 'poles' of the tube structure due to formation irregularities in the nanogratings as well as a different Young's modulus along the $y$-direction of the crystal (when compared to the $x$-direction) [42]. If this case is compared to that shown for 90° polarization (h), we find that the angles at locations 2 and 5 are smaller still, which may be correlated to the orientation of the writing polarization and the maximum anisotropic stress component. Similarly, the plots for 45° and 0° (f-g) show similar trends with regard to angle location and stress orientation. Note that, for the 0° case (c), the pattern is not distorted perfectly along the $x$-direction, due to slight misalignment of the sample with respect to the motion stages during fabrication.

Up to this point, measured retardance images have been used only for qualitative analysis, enabling the extraction of details about the formation and orientation of self-organized nanostructure as a by-product of the laser-writing process. For crystalline materials however, this may be taken a step further. Using simple image processing techniques, quantitative information may be extracted to yield information about the deformations present in the crystal lattice.

The retardance images shown in Fig. 6 were processed using contour plots, as shown in Fig. 7, producing a topological map for each pattern (a-d). Using the largest (lowest-level) contour, shown in dark blue, information about the angles between each facet of the contour was extracted using linear fitting, and plotted versus angle location, as shown in (e-h), with the representative locations shown in (i). For reference, the structure of the β-quartz is shown in (j).

At first glance, the apparent distortion in the retardance patterns is not clearly visible, and looks as though little (if any) contribution is present. However, when plotted with contours as shown in (a-d), the differences become immediately clear. The distortion is most evident when comparing the patterns for 45° (b) and 0° (c) case. It is worth nothing that the 0° degree pattern does not display symmetric distortion in relation to the polarization orientation (horizontally, along the bottom axis of the figure) as would be expected, and somewhat resembles the 45° pattern. This may be attributed to a slight misalignment polarization orientation with respect to of the substrate.

As with previous tests in fused silica, the undistorted retardance map for radial polarization (d) serves as a reference, against which linear polarization distortions may be gauged. This is reflected in the plot shown in (h), where the angles between each face are approximately equal (120±10°), as is expected for β-quartz. A few subtle differences are present, particularly at angle locations 2 and 5, which are smaller on average than the other four locations. This may be attributed to two factors: Slight irregularities in the nanograting formation at the 'poles' of the tube (see, for example Fig. 5a, radial polarization), and differences in the Young's modulus for $x$ and $y$ directions in the crystal lattice [42]. If we compare these results to the plot shown for 90° polarization (e), a similar trend is found. In this case however, locations 2 and 5 are even smaller, corresponding to the orientation of the maximum anisotropic stress component. Similarly, the plots for 45° and 0° (f-g) show similar trends.

## 4. Discussion and modeling

To explain the occurrence of the lobes that are characteristic of anisotropic retardance patterns generated by tube structures, we propose the simple model displayed in Fig. 8a. This model represents one quarter of a typical structure, written with fixed linear polarization oriented along the $y$-axis of the figure. The inset in (a) shows a magnified view, illustrating the orientation of the nanogratings and their role in stress generation, as highlighted by a single lamella. Experimental work by *Champion et al* has shown that nanogratings are responsible for a deformation on the order of ~0.03% [27], and that this deformation is largest in the direction perpendicular to the grating lamella (parallel to the beam polarization when using linear polarization). The profile of deformation (and hence the stress profile) is estimated by computing the chord length between the inner and outer radius of the ring, parallel to the $y$-axis, as a function of angle, resulting in the curve shown in red. This curve represents the deformation generated at the interface of the *outer* ring and the bulk substrate. The characteristic peak signifies a discontinuity, found when the angularly dependent $x$-position along the outer radius coincides with the maximum value of the inner radius. Here we have shown only the deformation profile in the $y$-direction; however a similar profile exists for the $x$-direction, albeit smaller in magnitude.

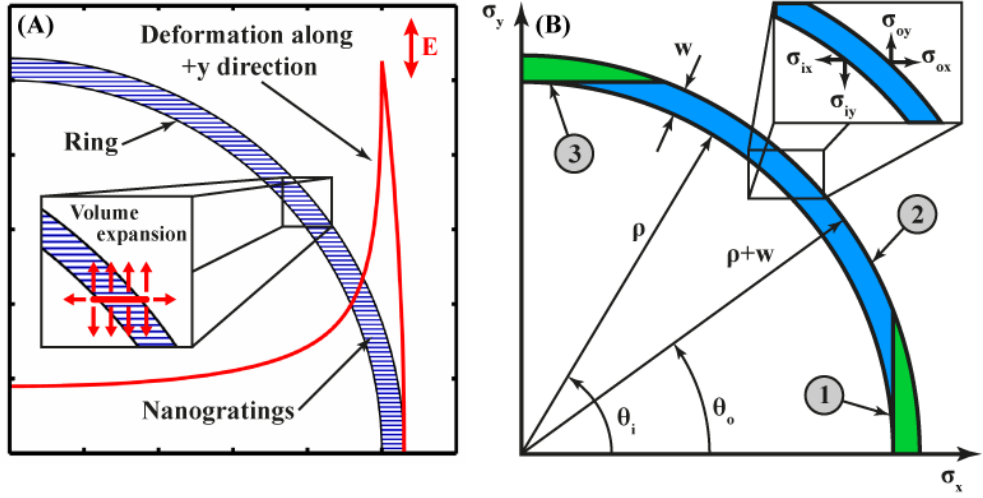

Fig. 8. Proposed model describing anisotropic stress loading in tube structures resulting from nanograting formation when using a linearly polarized writing beam. In this model, each grating lamella contributes to the overall deformation of the ring structure, as shown in (b), where only deformations along the $+y$ axis are considered for simplicity. The deformation curve, shown in red, may be calculated by taking the chord length along the $+y$ direction (perpendicular to the grating lamella), between the inner and outer radius, with respect to distance along the $+x$ direction. This model is further developed in (c) for both principal directions. Here the values of $\sigma_i$ and $\sigma_o$ represent the inner and outer stress profiles respectively. These values are computed with respect to the radius of the tube $\rho$, and the thickness of the laser modified zone $w$, as a function of angle around the tube radius. The indicated zones (1, 2, and 3) represent geometrically discontinuous regions which must be taken into account when calculating the stress profile for the outer radius, $\sigma_o$.

Tubes under stress represent a deformed system in static equilibrium. As such, forces acting on each surface of the structure must be accounted for, specifically those acting on the inner and outer radius. In Fig. 8b we further develop this model for chord lengths along the $x$ and $y$ directions, resolving stress components for the inner ($\sigma_i$) and outer radius ($\sigma_o$). The equations for the outer radius are as follows:

$$\sigma_{ox} \cong \begin{cases} \theta_o \leq \theta_{max} & \dfrac{w\cos\theta_o + \rho\sin\theta_o \left[\sin^{-1}\left\{\left(\dfrac{\rho+w}{\rho}\right)\sin\theta_o\right\} - \theta_o\right]}{\eta_{Lxo}} \\ \theta_o \geq \theta_{max} & \dfrac{(\rho+w)\cos\theta_o}{\eta_{Lxo}} \end{cases} \qquad (4)$$

$$\sigma_{oy} \cong \begin{cases} \theta_o \leq \theta_{min} & \dfrac{(\rho+w)\sin\theta_o}{\eta_{Lyo}} \\ \theta_o \geq \theta_{min} & \dfrac{w\sin\theta_o + \rho\cos\theta_o\left[\theta_o - \cos^{-1}\left\{\left(\dfrac{\rho+w}{\rho}\right)\cos\theta_o\right\}\right]}{\eta_{Lyo}} \end{cases} \qquad (5)$$

where $w$ is the thickness of the ring (laser-modified zone), and $\rho$ is the inner radius, while $\theta_i$ and $\theta_o$ are the angles along the inner and outer radius respectively. In practice, the stresses acting on the inner and outer radius are calculated simultaneously using the same angle. For the purposes of simulation, these functions have been normalized, with the normalization factors ($\eta_{Lxo}$, $\eta_{Lyo}$) given by:

$$\eta_{Lxo} = (\rho + w)\cos\theta_{max} \tag{6}$$

$$\eta_{Lyo} = (\rho + w)\sin\theta_{min} \tag{7}$$

The values of $\theta_{min}$ and $\theta_{max}$ represent the angular position (along the outer radius) of the discontinuity boundary for the y and x stress components respectively, and are calculated by:

$$\theta_{min} = \cos^{-1}\left(\frac{\rho}{\rho + w}\right) \tag{8}$$

$$\theta_{max} = \sin^{-1}\left(\frac{\rho}{\rho + w}\right) \tag{9}$$

Similar expressions are derived for the stress profile around the inner radius, as follows:

$$\sigma_{ix} \cong \frac{w\cos\theta_i + (\rho + w)\sin\theta_i \left[\theta_i - \sin^{-1}\left\{\left(\frac{\rho}{\rho + w}\right)\sin\theta_i\right\}\right]}{\eta_{Lxi}} \tag{10}$$

$$\sigma_{iy} \cong \frac{w\sin\theta_i + (\rho + w)\cos\theta_i \left[\cos^{-1}\left\{\left(\frac{\rho}{\rho + w}\right)\cos\theta_i\right\} - \theta_i\right]}{\eta_{Lyi}} \tag{11}$$

The normalization factors $\eta_{Lxi}$ and $\eta_{Lyi}$ are given by:

$$\eta_{Lxi} = \eta_{Lyi} = (\rho + w)\cos^{-1}\left(\frac{\rho}{\rho + w}\right) \tag{12}$$

This model was verified using a commercial finite-element-analysis software package (COMSOL) for a tube with a diameter of 100 µm ($\rho$=48 µm) and a (laser-modified) zone width $w$, of 2 µm. The simulation was verified using a 2D model, which was treated as a plane-strain problem due to the high aspect ratio of a typical tube structure. In this case, working from the findings of *Champion et al* [29], the peak stress was scaled to a maximum of 2 GPa, and a ratio between the y and x stress components of ~2.5. The results of this verification are shown in Fig. 9.

For comparison, Fig. 9 includes a representative retardance map (a), which has been scaled according to equation (3) to show difference principal stress. It is important to remember that the retardance values measured *within* the laser-modified zone are *not representative* of the generated stress in this region, and instead give an indication of the amount of form birefringence present in the self-organized nanostructure [15]. We have chosen the largest part of the 'lobes' (colored light green/blue in (a)) as a reference for comparison. These regions have a peak calculated stress of ~20 MPa, and are in good agreement with the stress map shown in (b).

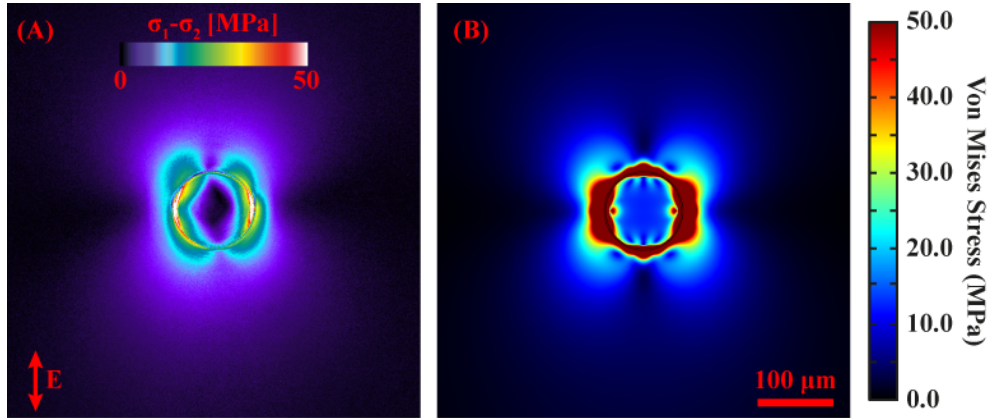

Fig. 9. Comparison between the stress calculated from a measured retardance pattern (a) and a simulated tube structure with a diameter of 100 μm (b). The simulated peak stress was scaled to a value of 2 GPa, with a ratio between the *y* and *x* components of the stress of ~2.5 [27].

Overall, the general shape and distribution of the stress displayed in the model matches well with that of the original structure, however some subtle differences are immediately apparent. In the 'equatorial' regions of the simulated structure, we find that the stress concentrations are larger and broader than those in the original tube structure. These inaccuracies may possibly be explained through our treatment of the laser-modified zone in the model. In reality, this region is heavily modified by the laser, and may not be assumed to have the same mechanical properties as the original un-modified substrate. Since new material properties are currently unknown, we have chosen to represent this region with the same material as the substrate. The ring geometry is therefore only used to define the stress profile around the laser-written structure.

## 5. Conclusion

Understanding anisotropy of laser-induced material modifications is fundamentally important to the process of laser machining. The effects on device manufacture and performance can be extreme: either contributing negatively or as a benefit.

In this work, we introduced a simple procedure to assess these effects. Using photo-elastic measurements of symmetric laser-written patterns, we provide a quantifiable and efficient means of *direct* observation of stress-anisotropy. These structures induce a localized stress field – and consequently an optical retardance pattern – whose size, shape, and intensity gives an indication of the origins of anisotropy in the laser writing process.

As an illustration, we applied this approach to fused silica and quartz, which are amorphous and crystalline phases of the same material ($SiO_2$).

In fused silica, using this method, we emphasized the importance of polarization as it relates to stress generation. In particular, we examined unconventional symmetric polarization states such as radial and azimuthal. Using a simple model based on volume expansion, we confirmed the link between nanograting orientation and stress intensity, and showed how this model can be used a predictive tool. Our model suggests that stresses as high as 2 GPa are found in the vicinity of the nanograting lamellae.

In a crystalline phase of silica (β-quartz), we used our method to detect the presence of added stress-anisotropy, induced as a result of the laser-writing process, which we linked to the presence of nanogratings. Incidentally, our method can also be used for finding crystal orientation in unknown specimens.

Although we have only demonstrated the method for two representative materials, it remains generally applicable to a wide array of transparent dielectrics, whether they are amorphous or crystalline.

**Acknowledgements**

The authors acknowledge the financial support from the European Research Council under the Galatea Project (ERC-2012-StG-307442), http://www.erc-galatea.eu. The authors also acknowledge Prof. Peter Kazansky from the University of Southampton for preparing the passive polarization polarization converter used in this work, part of the Femtorpint project, http://www.femtoprint.eu.